# Moiré Modulated Lattice Strain and Thickness-Dependent Lattice Expansion in Epitaxial Ultrathin Films of PdTe$_2$

Jacob Cook,[1] Dorri Halbertal,[2] Qiangsheng Lu,[1] Mathew Snyder,[1] Yew San Hor,[3] Dmitri N. Basov,[2] Guang Bian[1*]

*Correspondence and requests for materials should be addressed to G.B. (E-mail: biang@missouri.edu)

[1]Department of Physics and Astronomy, University of Missouri, Columbia, Missouri 65211, USA

[2]Department of Physics, Columbia University, New York City, New York 10027, USA

[3]Department of Physics and Astronomy, Missouri University of Science and Technology, Rolla, Missouri 65409, USA

**Abstract**

We report the epitaxial growth of PdTe$_2$ ultrathin films on topological insulator Bi$_2$Se$_3$. A prominent moiré pattern was observed in STM measurements. The moiré periodicity increases as film thickness decreases, indicating a lattice expansion of epitaxial PdTe$_2$ thin films with lower thicknesses. In addition, our simulations based on a multilayer relaxation technique reveal uniaxial lattice strains at the edge of PdTe$_2$ domains, and anisotropic strain distributions throughout the moiré supercell with a net change in lattice strain up to ~2.9%. Our DFT calculations show that this strain effect leads to a narrowing of the band gap at Γ point near the Fermi level. Under a strain of ~2.8%, the band gap at Γ closes completely. Further increasing the lattice strain makes the band gap reopen and the order of conduction band and valence bands inverted in energy. The results offer a proof of concept for constructing quantum grids of topological materials under the modulation of moiré potentials.

**Introduction**

One of the most extensively sought-after phenomena in condensed matter is topological superconductivity (TSC) due to the possibility of such novel electronic states hosting Majorana fermions, which arise from the spin-triplet, *p*-wave superconductivity, and surface electronic states [1,2]. These exotic quasiparticles are of significant interest for their application in the field of quantum computing, due to the topologically protected nature which is robust against decoherence [3-8]. A promising route to realizing topological superconductivity is inducing superconductivity in topological insulators via proximity effect with a superconductor [9]. The topological superconducting heterostructure can be constructed via epitaxial growth of a superconductor film on the surface of a topological insulator [9,10]. It is desirable to use two van der Waals (vdW) materials as components of the heterostructure for an atomically sharp interface that can reduce the interfacial disorders and enhance the proximity effects.

One well studied and robust topological insulator is $Bi_2Se_3$, which can be cleaved in ultra-high vacuum to produce a pristine surface to use as a substrate. The transition metal dichalcogenide (TMDC) semimetal $PdTe_2$ has been known to be a bulk superconductor with a critical temperature of 1.7K. In addition, the in-plane crystal lattice parameters of $PdTe_2$ are nearly commensurate to those of $Bi_2Se_3$. $PdTe_2$ is also known to host rich electronic properties such as intrinsic type-I superconductivity [11-15] with anomalous type-II surface behavior [16,17], a Type-II Dirac cone [18-21], topologically non-trivial surface states [18,22,23], and a semiconductor to conductor transition from 1 to 2 monolayers (ML) [24]. More importantly, it has been shown that $PdTe_2$ maintains its superconducting state even with thickness down to two monolayers (ML) and $T_c$ = 1.42 K at 2ML [24]. Both $PdTe_2$ and $Bi_2Se_3$ are VdW materials, which is ideal for constructing a heterostructure with a smooth interface. Therefore, building a heterostructure of $Bi_2Se_3$ and $PdTe_2$ offers an opportunity to study the interaction of superconductivity with topological Dirac states.

An important phenomenon pertinent to ultra-thin films and heterostructures is moiré patterns, which can arise from twisting rotations between adjacent crystal layers [25] and the lattice mismatch at the interface of two films [26,27]. The moiré patterns and related long-range potentials have profound effects on the electronic and magnetic properties of heterostructures, such as superconductivity of twisted bilayer graphene [25], modulation of edge modes [27], and tunable electronic properties [27-31].

In this work, we perform the epitaxial growth of $PdTe_2$ superconducting films on topological insulator $Bi_2Se_3$. The electronic structure of heterostructure is probed by angle-resolved photoemission spectroscopy (ARPES). Our scanning tunneling microscope (STM) measurements reveal a very strong and dynamic moiré pattern that arises from the lattice mismatch between the $PdTe_2$ and $Bi_2Se_3$ substrate. A thickness dependency of the moiré pattern is observed, suggesting increases in moiré periodicity with decreasing thickness. Theoretical analysis of the moiré system reveals inhomogeneous lattice strains in $PdTe_2$ films, and a band inversion induced by lattice strains.

**Materials and methods**

The $PdTe_2$ thin films were grown on a $Bi_2Se_3$ substrate in an integrated MBE-STM ultrahigh vacuum (UHV) system with base pressure below $2\times10^{-10}$ mbar. The $Bi_2Se_3$ was prepared by *in-situ* cleaving the surface and subsequent annealing to 250°C for 1 hour. Then, high-purity Pd (99.95%) and Te (99.9999%) were evaporated from an electron-beam evaporator and a standard Knudsen cell, respectively, with a flux ratio of 1:10. The deposition rate of Pd and Te atoms was monitored by a quartz oscillator. The temperature of substrate was kept at 210°C during the growth. The surface topography mapping was performed *in-situ* by using the SPECS Aarhus-150 STM with a Tungsten tip. After STM, the film was transferred *in-situ* to the ARPES stage. ARPES measurements were performed at 100 K using a SPECS PHOIBOS-150 hemisphere analyzer with a SPECS UVS 300 helium discharge lamp (He I$\alpha$ = 21.2 eV and He II$\alpha$ = 40.8 eV). The energy resolution is 40 meV at 100 K.

Thickness dependent lattice expansion and relaxation was calculated using a multilayer relaxation technique [32], a newer methodology for stacked thin film crystals utilizing material parameters, sheer moduli, and generalized stacking fault energy to calculate interactions and relaxations of layered materials.

We performed first-principles calculations with density functional theory as implemented in the Vienna *ab initio* Simulation Package (VASP). We used the Perdew-Burke-Ernzerhof (PBE) form for the exchange-correlation functional with a plane-wave cut-off energy of 300 eV. The super cell includes $PdTe_2$ layers and a vacuum layer of about 20 Å. The $PdTe_2$ layers were allowed to relax during the geometry optimization.

**Results and Discussion**

PdTe$_2$ has a layered trigonal crystal structure with a unit cell of a hexagonal Pd layer sandwiched between Te layers. Bulk PdTe$_2$ has the 1T-CdI$_2$ structure type and a *P*3*m*1 space group with AA stacking, meaning there is no rotation between layers. The bulk lattice parameters are a = 4.028Å in-plane for PdTe$_2$ [24] and a = 4.14Å for Bi$_2$Se$_3$ [33] [Fig. 1(a)] in the literature. The STM topography of the growth [Fig. 1(b)] shows high quality PdTe$_2$ crystals on the Bi$_2$Se$_3$ surface. The edge angles of PdTe$_2$ islands are mostly 120°, reflecting the hexagonal atomic lattice. A honeycomb-like moiré pattern is visible in PdTe$_2$ films with different thicknesses. Atomic-resolution STM images of the Bi$_2$Se$_3$ substrate and bulk PdTe$_2$ growth [Fig. 1(c,d)] show the hexagonal lattice structure with the average lattice constants of a = 4.15±0.1Å and a = 3.98±0.1Å for each film, respectively. These measurements match well with the values in the literature. ARPES measurements were used to confirm the high crystallinity of both the substrate and epitaxial films [Fig. 1(e,f)]. The ARPES band structure of Bi$_2$Se$_3$ [Fig. 1(e)] clearly shows the band gap between the conduction and valence bands and the topologically protected surface states (TSS). The ARPES results of 10 ML PdTe$_2$ [Fig. 1(f)] is consistent with the ARPES result of pristine PdTe$_2$ films and band structure calculations (See Supplementary Information).

To understand the moiré pattern observed in the PdTe$_2$/Bi$_2$Se$_3$ heterostructure, we need to determine the relation between the lattice parameters of the two surfaces. Lattice parameters play an important role in how the crystal grows on the substrate. Lattices with close to commensurate conditions are likely to align with minimal rotations. Meanwhile, a slight difference in the magnitudes of lattice parameters can cause a long-range modulation in the alignment of the interfacial atoms between the two crystals. If there are sufficient interfacial interactions, then the change in atomic alignment at the interface can lead to a measurable, periodic pattern on the crystal surface. By taking the interfacial atoms of the PdTe$_2$/Bi$_2$Se$_3$ heterostructure and overlaying them [Fig. 2(a)], we can see this long-range moiré pattern form from the alignment of the top Bi and Se atoms of the substrate and the bottom Pd and Te of the PdTe$_2$. This reveals three primary stacking configuration regions at the interface [Fig. 2(a)]: (1) sites where the Bi and Te atoms align and the Se and Pd atoms align, (2) sites where the Se and Te atoms align, and (3) sites where the Bi and Pd atoms align, as marked by letters "B-D" in Fig. 2(a). To find the moiré periodicity, we simply find the distance between the regions with same stacking order. We can calculate the moiré pattern periodicity by aligning the PdTe$_2$ and Bi$_2$Se$_3$ crystal lattices with a 0° twisting angle and finding the distance required to realign the atomic positions on the two surfaces. Using this method with the lattice parameter in the literature, the moiré periodicity is found to be 14.49 nm. The proximity of the interfacial atoms in the different alignment regions will affect the stacking energy and the local density of states (LDOS) and can lead to variations in the transport properties along the surface. Using the multilayer relaxation technique [32] with the generalized stacking fault energy (GSFE), the stacking energy density (SED) was calculated across films with thickness ranging from 3 ML to 12 ML. The calculated SED of 3 ML and 10 ML $PdTe_2$ on a rigid $Bi_2Se_3$ substrate is shown in Figs. 2(e,f). The map of the SED exhibits a honeycomb-like moiré pattern for 3ML and a more hexagonal moiré pattern (more prominent in large-scale STM image in Fig. 3) for 10ML. The transition from the honeycomb-like to the hexagonal moiré is gradual as the films thickness increases.

The moiré pattern can be directly resolved by STM since the moiré potential gives rise to a change in the LDOS at the interface. For the 10ML sample [Fig. 3(a)], the moiré pattern has the hexagonal structure with a periodicity of 14.255 nm, slightly smaller than the theoretical prediction. This indicates the lattice constants of 10ML $PdTe_2$ is slightly smaller than the literature values. In contrast to the hexagonal moiré pattern of the 10ML sample, the sample of 3-5 ML [Fig. 3(b)] demonstrate a honeycomb-like pattern. The observed periodicity of the moiré pattern in the 3ML region is a 20.114 nm, significantly larger than the literature value, indicating a pronounced tensile lattice strain in 3 ML. The moiré pattern is continuous across the edges of 4ML and 5ML patches, confirming that the pattern itself arises from the interfacial coupling between the $Bi_2Se_3$ substrate and the $PdTe_2$ thin film, rather than a rotation between layers of the $PdTe_2$. By examining the moiré pattern along difference directions and on different samples, the averaged moiré periodicity for the varying thicknesses of $PdTe_2$ films was obtained. The thickness dependence of the averaged moiré periodicity [Fig. 3(a)] shows that moiré periodicity increases monotonically as the thickness of $PdTe_2$ films decreases. In our calculations, the lattice constants of the $Bi_2Se_3$ substrate are fixed since it is a bulk crystal with a much more rigid structure. The calculated lattice constants of $PdTe_2$ films are plotted in Fig. 3(d). It shows that there is a thickness-dependent lattice expansion as the film thickness decreases, which is consistent with theory. The average lattice expansion of the 3ML $PdTe_2$ film is about 0.7%. We note that the calculated lattice parameter is a global value averaged over many samples and thus it shows no information about the local variations in the lattice expansion.

Interestingly, the moiré pattern can be distorted by the local lattice irregularities such as edges and defects [Fig. 4]. Near the edge of the 3ML sample, we observed a region with uniaxial distortion of moiré pattern [Fig. 4(a)]. To calculate the SED and relaxation for non-periodic structures, a modified version of the multilayer relaxation code was designed for the non-trivial strain cases. Under this methodology, the film was treated as a single membrane. The uniaxial distortion is in good agreement with the SED calculated by using this theory [Fig. 4(b)] [32]. The SED map is obtained by minimizing the surface energy near the edge. The compressive strain map for the 3ML edge data [Fig. 4(c)] reveals that the uniaxial compression of $PdTe_2$ lattice increases as one moves closer to the edge of the crystal. This can be quantified by plotting the moiré periodicity as a function of the distance in both perpendicular ($\alpha$) and parallel ($\gamma$) directions relative to the edge [Fig. 4(d)]. The uniaxial distortion of moiré pattern towards to the edge is due to the interplay of the interfacial effect and the breaking of the translational symmetry at the edge. The moiré periodicity has little changes parallel to the edge because the translational symmetry is intact in the $\gamma$ direction. We also observed an irregular distortion of moiré pattern in the central region of the 3ML sample [Fig. 4(e)]. This anisotropy in the distorted moiré pattern can be reproduced by plotting the SED over the distorted region [Fig. 4(f)]. From the theoretical model, we extracted the compressive and sheer strain in $PdTe_2$ lattice at both the uniform and distorted regions. Near the distorted region, the strain in $PdTe_2$ lattice ranges from a compression of -0.7% at the high LDOS regions to a massive 2.2% expansion at the centers of the honeycomb moiré supercell [Fig. 4(g)]. This corresponds a net change of 2.9% in the lattice parameter of $PdTe_2$ at different locations of moiré supercells. A zoom-in image shows the local variations of lattice strains within a single moiré supper cell [Fig. 4(h)]. This net strain of 2.9% is large enough to make the $PdTe_2$ lattice parameter perfectly match the $Bi_2Se_3$ substrate.

The large local lattice strain can have significant effects on the electronic band structure of $PdTe_2$ films. We calculated the electronic structure with varied lattice constants under the framework of the density functional theory (DFT). The DFT calculations show an interesting change in the band structure of 3ML $PdTe_2$ near the $\Gamma$ point at the Fermi level ($E_f$) [Figs. 4(i-l)]. With a strain of -0.7%, which corresponds to a slight lattice compression, there is an inverted band gap between the $p_z$ and $p_{x,y}$ bands with a gap size of 0.125eV [Fig. 4(i)]. At the observed maximum lattice expansion of 2.2%, the band gap shrinks to 0.055eV [Fig. 4(j)]. The band gap closes completely at the expansion of 2.8% [Fig. 4(k)]. This theoretical compressive strain accidentally matches with the maximum possible lattice strain of $PdTe_2$ on $Bi2Se_3$, at which the two lattice constants of PdTe$_2$ on Bi$_2$Se$_3$ are identical. Further increasing the lattice expansion leads to reopening of the band gap and restoring the normal order of the $p_z$ and $p_{x,y}$ bands in energy [Fig. 4(i)]. This shows that under sufficient strains with a moiré supercell, 3ML PdTe$_2$ can undergo a topological transition from topological insulators to trivial insulators. Interestingly, this topological phase transition is localized and confined to each moiré supercell. In other words, within each moiré supercell, there exists a core region of topological insulator with topologically protected surface states. This is literally equivalent to a grid of topological insulator quantum dots with size of tens of nanometers. The separation between topological insulator quantum dots is controlled by the moiré periodicity of heterostructure. Though the observed lattice strain in the PdTe$_2$/Bi$_2$Se$_3$ heterostructure is insufficient to induce the band inversion, our results offer an intriguing proof of concept for realizing a moiré patterned grid of topological insulator quantum dots.

**Conclusion**

We grew PdTe$_2$ thin films with thickness down to 3ML on topological insulator Bi$_2$Se$_3$ by using MBE. A very prominent moiré pattern was observed in the STM mapping. The moiré periodicity increases as film thickness decreases, indicating a lattice expansion of PdTe$_2$ epitaxial thin films. This moiré pattern arises from the interface lattice mismatch between the PdTe$_2$ and Bi$_2$Se$_3$ crystals. The observed moiré pattern shows the in-plane lattice parameter PdTe$_2$ epitaxial films approaches that of Bi$_2$Se$_3$ in the 2D limit. The SED and in-plane strain maps for this heterostructure reveal uniaxial strain at the edge and anisotropic strain distributions throughout the moiré supercell with a net change in lattice strain of PdTe$_2$ up to ~2.9%. Our DFT calculations show that this strain effect leads to a narrowing of the band gap at Γ point near the Fermi level. Under a strain of ~2.8%, the maximum possible lattice strain at which PdTe$_2$ becomes commensurate with Bi$_2$Se$_3$, the band gap closes completely. Further increasing the lattice strain (for example, by using a substrate with a larger lattice constant compared to Bi$_2$Se$_3$) makes the band gap reopen and the order of conduction band and valence bands inverted in energy. This means that under enhanced strain, the band topology of PdTe$_2$ undergoes a phase transition. Furthermore, since the maximum lattice strain is observed at the center of the moiré supercell, the topological phase with nontrivial band order is localized and confined in each moiré supercell in real space. This opens a new avenue of producing topological phase transitions in real space and creating a nanometer-sized grid of topological insulators for device applications.

## Acknowledgments

Research on atomic relaxation at Columbia is supported by W911NF2120147. DNB is Moore Investigator in Quantum Materials EPIQS GBMF9455. DH was supported by a grant from the Simons Foundation (579913).

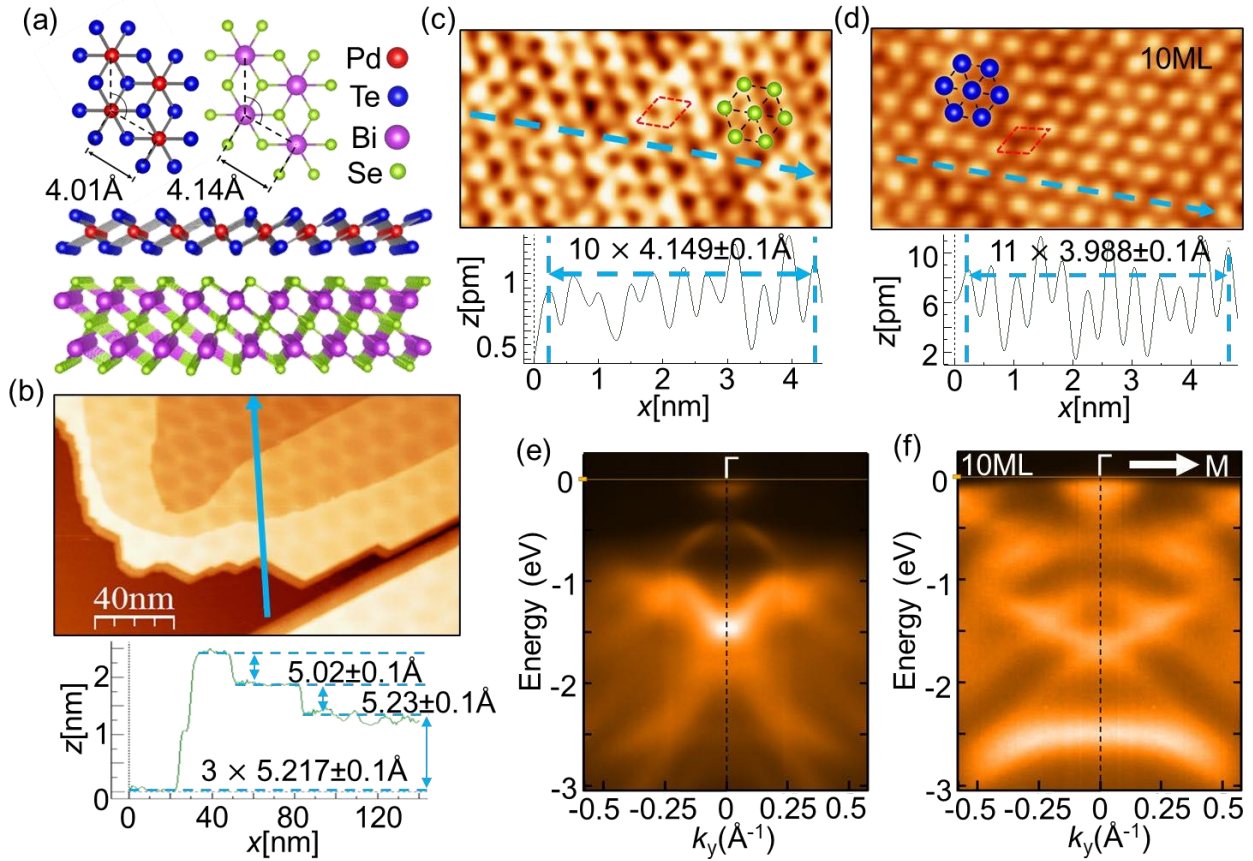

**Figure 1.** Crystal structures, STM, and bulk ARPES measurements. (a) Lattice parameters and crystal structure of PdTe$_2$ crystals, shown with a Bi$_2$Se$_3$ layer beneath. (b) STM topography of PdTe$_2$ growths on Bi$_2$Se$_3$ substrate. Inset shows the height profile designated by the blue arrow. The totally step height is 1.638nm, slightly larger than three times the c = 5.118Å out of plane lattice constant. (c) Atomic-resolution STM of the Bi$_2$Se$_3$ substrate and (d) 10ML PdTe$_2$ crystal surface, both showing a hexagonal lattice from the surface atoms as shown in the overlay. Periodicity was determined by a cut along the blue arrow, which is graphed below each image. (e-f) ARPES data for the Bi$_2$Se$_3$ substrate (e) and bulk PdTe$_2$ growth (f), both showing good quality crystals.

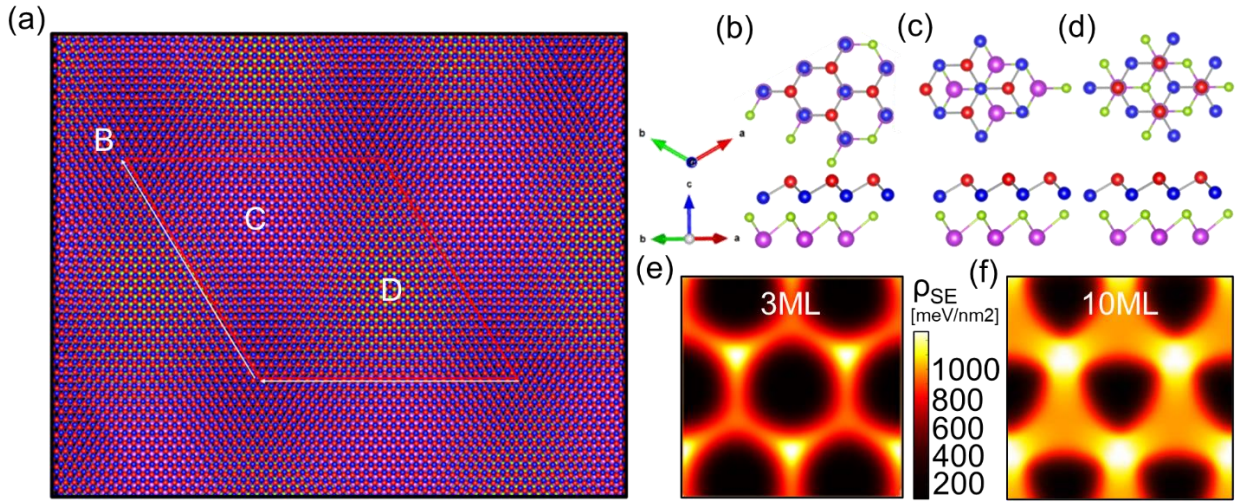

**Figure 2.** Moiré pattern of PdTe$_2$/Bi$_2$Se$_3$ heterostructure. (a) Overlay of PdTe$_2$ on top of Bi$_2$Se$_3$, revealing three primary alignments between PdTe$_2$ and Bi$_2$Se$_3$, denoted as B-D. (b) AA' stacking of PdTe$_2$ on Bi$_2$Se$_3$ corresponding to region 'B' in (a). (c) A'B stacking of PdTe$_2$ on Bi$_2$Se$_3$ corresponding to region 'C' in (a). (d) AB' stacking of PdTe$_2$ on Bi$_2$Se$_3$. corresponding to region 'D' in (a). (e) The calculated SED for a 3ML PdTe$_2$ film on a rigid Bi$_2$Se$_3$ substrate, revealing the honeycomb-like moiré pattern. (f) The calculated SED for a 10ML PdTe$_2$ film on a rigid Bi$_2$Se$_3$ substrate, showing the transition to a hexagonal moiré pattern.

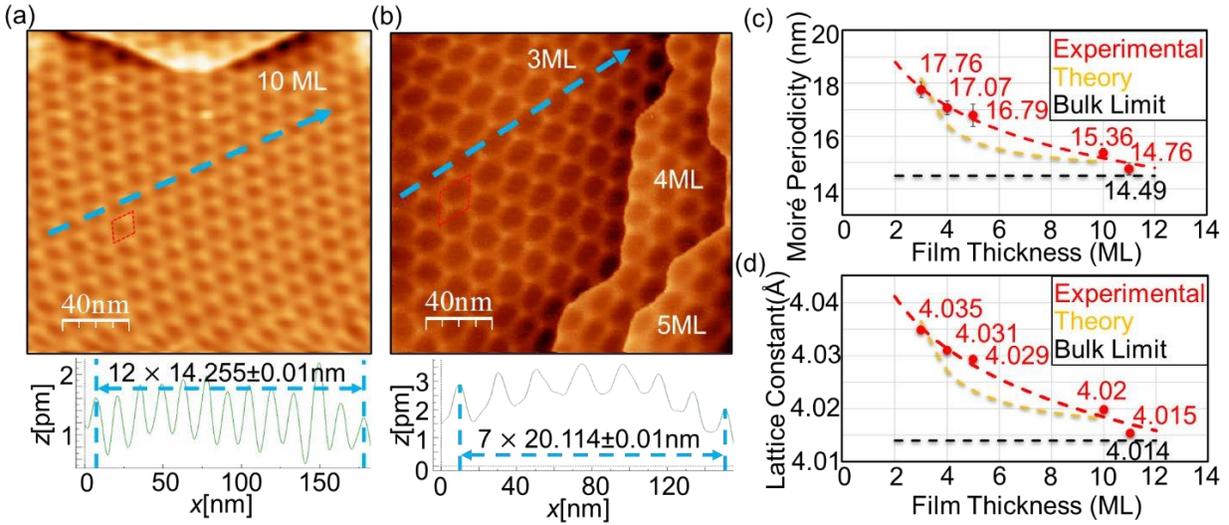

**Figure 3.** Moiré Periodicity and Thickness-Dependent Lattice Expansion. (a) STM image of 10ML PdTe$_2$ with the hexagonal moiré superlattice. The moiré periodicity is shown in the height profile below. (b) STM image of 3ML PdTe$_2$ with the honeycomb moiré superlattice. The moiré periodicity is shown in the height profile below. (c) Thickness dependance of the moiré periodicity averaged over multiple samples and compared with theoretical predictions. (d) Thickness dependance of the averaged lattice constant derived from the experimental and theoretical moiré pattern.

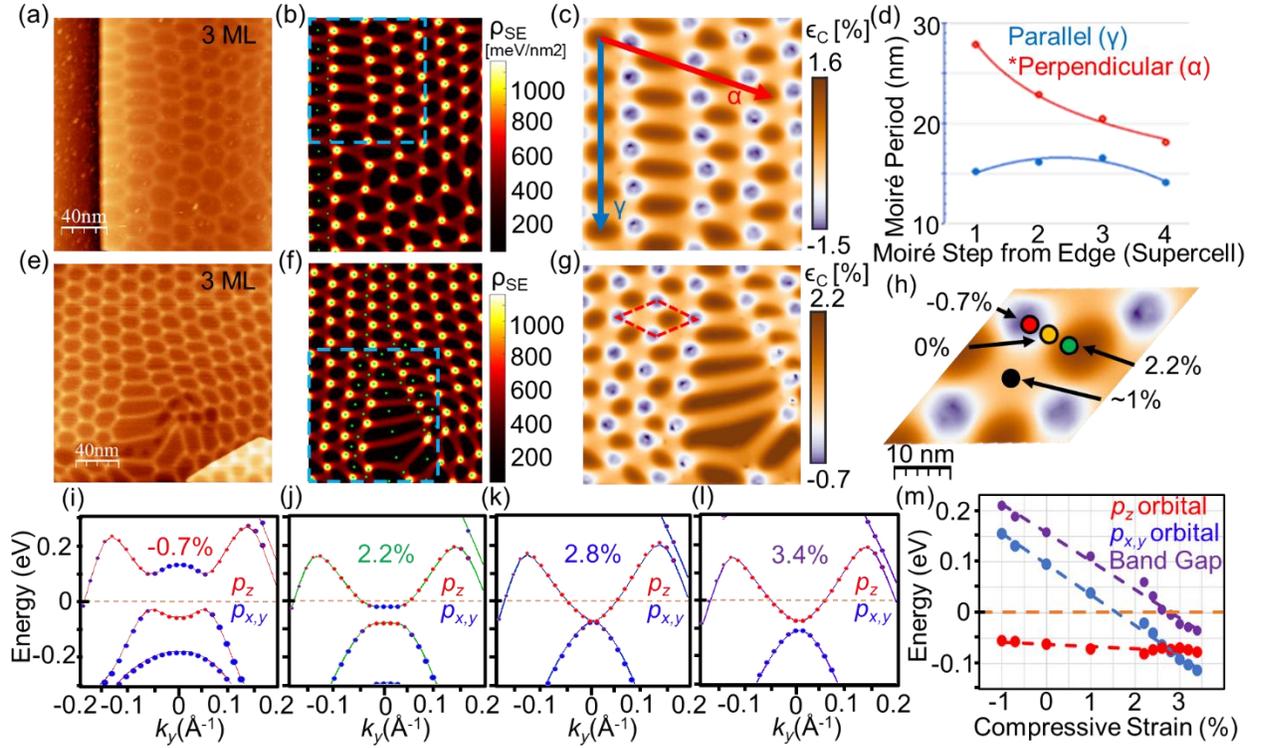

**Figure 4.** Anisotropic Moiré and Strain Analysis. (a) STM image of 3ML PdTe$_2$ with an edge distortion of moiré pattern. (b) SED mapping of the STM data in (a). The green dots correspond to points held fixed during relaxation. (c) In-plane compressive strain map of the isotropic edge inscribed by the blue square in (b), showing the moiré periods in the α and γ directions. (d) Uniaxial edge strain effect, enhancing strain perpendicular to the edge compared to the parallel direction. (e) STM image of 3ML PdTe$_2$ with an inhomogeneous moiré region. (f) SED mapping of the STM data in (e). The green dots correspond to points held fixed during relaxation. (g) In-plane compressive strain map of the anisotropic region inscribed by the blue square in (f) with a net strain magnitude from -0.7% to 2.2%. (h) Close-up moiré Supercell from (g) with denoted localized strain magnitudes. (i-l) Calculated electronic structure with orbital components near E$_f$ and Γ point of 3ML PdTe$_2$ with compressive strain values ranging from -0.7% to 3.4%. (m) Calculated upper and lower band binding energies with respect to changing strain, revealing a process of band inversion at Γ point. The band inversion occurs at the strain of 2.81%.

# References

[1] M. Sato and Y. Ando, Rep. Prog. Phys. **80**, 076501 (2017).

[2] L. P. Gor'kov and E. I. Rashba, Physical Review Letters **87**, 037004 (2001).

[3] D. Litinski, M. S. Kesselring, J. Eisert, and F. von Oppen, Physical Review X **7**, 031048 (2017).

[4] X.-L. Qi and S.-C. Zhang, Reviews of Modern Physics **83**, 1057 (2011).

[5] S. D. Sarma, M. Freedman, and C. Nayak, npj Quantum Information **1**, 15001 (2015).

[6] J. L. Zhang, PNAS **108**, 24 (2011).

[7] P. Hosur, P. Ghaemi, R. S. K. Mong, and A. Vishwanath, Physical Review Letters **107**, 097001 (2011).

[8] C. Q. Han, Applied Physics Letters **107**, 171602 (2015).

[9] H. Sun *et al.*, Physical Review Letters 116 257003 (2016)

[10] F. Yang *et al.*, Physical Review B **86**, 134504 (2012).

[11] S. Das, Amit, A. Sirohi, L. Yadav, S. Gayen, Y. Singh, and G. Sheet, Physical Review B **97**, 014523 (2018).

[12] H. Leng, C. Paulsen, Y. K. Huang, and A. de Visser, Physical Review B **96**, 220506 (2017).

[13] H. Leng, J. C. Orain, A. Amato, Y. K. Huang, and A. de Visser, Physical Review B **100**, 224501 (2019).

[14] Amit and Y. Singh, Physical Review B **97**, 054515 (2018).

[15] P. Garcia-Campos, Y. K. Huang, A. de Visser, and K. Hasselbach, Physical Review B **103**, 104510 (2021).

[16] A. Sirohi, Journal of Physics: Condensed Matter **31** (2019).

[17] T. Le, L. Yin, Z. Feng, Q. Huang, L. Che, J. Li, Y. Shi, and X. Lu, Physical Review B **99**, 180504 (2019).

[18] F. Fei *et al.*, Physical Review B **96**, 041201 (2017).

[19] W. Zheng *et al.*, Physical Review B **97**, 235154 (2018).

[20] H.-J. Noh, J. Jeong, E.-J. Cho, K. Kim, B. I. Min, and B.-G. Park, Physical Review Letters **119**, 016401 (2017).

[21] S. Teknowijoyo, N. H. Jo, M. S. Scheurer, M. A. Tanatar, K. Cho, Physical Review B **98**, 024508 (2018).

[22] O. J. Clark *et al.*, Physical Review Letters **120**, 156401 (2018).

[23] Y. Lui, Chinese Review Letters **32** (2015).

[24] C. Liu *et al.*, Physical Review Materials **2**, 094001 (2018).

[25] G. Tarnopolsky, A. J. Kruchkov, and A. Vishwanath, Physical Review Letters **122**, 106405 (2019).

[26] B. A. Parkinson, F. S. Ohuchi, K. Ueno, and A. Koma, Applied Physics Letters **58**, 472 (1991).

[27] S. Kezilebieke, V. Vaňo, M. N. Huda, M. Aapro, S. C. Ganguli, P. Liljeroth, Nano Letters **22**, 328 (2022).

[28] E. Y. Andrei, D. K. Efetov, P. Jarillo-Herrero, A. H. MacDonald, Nature Reviews Materials **6**, 201 (2021).

[29] Y. Cao, D. Rodan-Legrain, O. Rubies-Bigorda, J. M. Park, K. Watanabe, and T. Taniguchi, Nature **583**, 215 (2020).

[30] D. M. Kennes *et al.*, Nature Physics **17**, 155 (2021).

[31] Z. Zheng *et al.*, Nature **588**, 71 (2020).

[32] D. Halbertal, L. Kleble, V. Hseih, J. Cook, S. Carr, G. Bian, C. Dean, D. Kennes, and D. Basov, arXiv:2206.06395 (2022)

[33] M. Xie *et al.*, Chinese Phys. B **22** 068101 (2013)